\documentclass[a4paper,10pt,twoside]{cpc-hepnp}

\usepackage{multicol}
\usepackage{graphicx}
\usepackage{booktabs}
\usepackage{amssymb,bm,mathrsfs,bbm,amscd}
\usepackage[tbtags]{amsmath}
\usepackage{lastpage}

\begin{document}

\fancyhead[co]{\footnotesize H. Sadeghi~ et al: The astrophysical
Structure of A = 11 double-$\Lambda$ hypernuclei studied ...}

\footnotetext[0]{Received -- March ---}

\title{The Astrophysical S-factor of the $^{12}$C($\alpha$,$\gamma$)$^{16}$O Reaction
at Solar Energies}

\author{%
      H. Sadeghi\email{H-Sadeghi@Araku.ac.ir}%
\quad R. Ghasemi \quad H. Khalili} \maketitle

\address{%
Department of Physics, Faculty of Science, Arak University, Arak 8349-8-38156, Iran\\
}

\begin{abstract}
The astrophysical S-factor of the $^4$He-$^{12}$C radiative capture
is calculated in the potential model at the energy range 0.1-2.0
MeV. Radiative capture $^{12}$C($\alpha$,$\gamma$)$^{16}$O is
extremely relevant for the fate of massive stars and determines if
the remnant of a supernova explosion becomes a black hole or a
neutron star. Because this reaction occurs at low-energies the
experimental measurements is very difficult and perhaps impossible.
In this paper, radiative capture of the
$^{12}$C($\alpha$,$\gamma$)$^{16}$O reaction  at very low-energies
is taken as a case study. In comparison with other theoretical
methods and available experimental data, good agreement is achieved
for the astrophysical S-factor of this process.
\end{abstract}

\begin{keyword}
Radiative capture, The astrophysical S-factor, Potential model
\end{keyword}

\begin{pacs}
25.55.-e,21.60.De,27.20.+n,26.20.Cd
\end{pacs}

\begin{multicols}{2}

\section{Introduction}

When the star's hydrogen-burning phase transition, star is helium
burning phase. The thermal energy at 1.5$\times$10$^8$ K  is
sufficient for the fusion of two helium nuclei Thus, the two helium
nuclei become unstable nuclei$^8$Be. If the conditions are suitable
Nuclei $^8$Be by the capture of the $\alpha$-particle radiation
convert to the Nuclei $^{12}$C. Process forms the nucleus is called
triple-$\alpha$ process \cite{Opik,Salpeter}.
\begin{equation}
^4He+^4He+^4He\leftrightarrow ^8Be+^4He\rightarrow ^{12}C+\gamma
\end{equation}
Therefore, conditions at  1.5$\times$10$^8$ K  is provided for
$\alpha$-particle capture by Nuclei$^{12}$C   and Nuclei $^{16}$O is
produced:
\begin{equation}
^{12}C+^4He\rightarrow ^{16}O+\gamma
\end{equation}
These reactions are importance due to carbon and oxygen are the most
abundant elements in the world after burning helium and often
heavier elements are formed from these two elements. Then Our
knowledge of the reaction $^{12}$C($\alpha$,$\gamma$)$^{16}$O
helpful to better understand the evolution of condense stars, such
as neutron stars and black holes. For example, a large cross section
for this reaction leads to the production of heavier elements. While
small cross section can lead to the reverse situation and production
of lighter elements. Thus, Our main purpose is to calculate cross
section of this process.

$^{12}$C($\alpha$,$\gamma$)$^{16}$O radiative capture process plays
a major role in stars fuel when they collapse. There is no accurate
and complete information about these reactions. Because the cross
section of this reaction is low and impossible to produce in the
laboratory directly at
low-energies~\cite{Eddington,Gamow,Atkinson,Rolfs,Burbidge}.

In past decades, the yield of capture rays has been studied for
$E_\alpha$ up to 42 MeV~\cite{Ajzenberg}. The cross sections of
$^{12}$C($\alpha$,$\gamma$)$^{16}$O capture process have been
obtained by fitting measured cross sections and extrapolating them
to low-energies utilizing standard R-matrix, Hybrid R-matrix and
K-matrix procedures.  The influence of vacuum polarization effects
on sub barrier fusion is also evaluated in~\cite{Assenbaum}, and the
relevance of Coulomb dissociation of $^{16}$O into $^{12}$C+$\alpha$
is studied in~\cite{Baur1,Baur2,Shoppa}. Calculations to test the
sensitivity of stellar nucleosynthesis to the level in $^{12}$C at
7.74 MeV are described in~\cite{Xiang}.

Recently, Dubovichenko et al., have been calculated the
astrophysical S-factor of the $^4$He--$^{12}$C radiative capture
using cluster model at the energy range 0.1-4.0 MeV. They shown that
the approach used, which takes into account E$_2$ transitions only,
gives a good description of the new experimental data for adjusted
parameters of potentials and leads to the value S(300)=16.0
keV.b~\cite{Dubovichenko1,Dubovichenko2}. More recently, Bertulani
presented a computer program aiming at the calculation of bound and
continuum state observables for a nuclear system such as, reduced
transition probabilities, phase-shifts, photo-disintegration cross
sections, radiative capture cross sections, and the astrophysical
S-factors~\cite{Bertulani}. The code is based on a potential model
type and can be used to calculate nuclear reaction rates in numerous
astrophysical reactions. In order to calculate the direct capture
cross sections one needs to solve the many-body problem for the
bound and continuum states of relevance for the capture process.  A
model based on potential can be applied to obtain single-particle
energies and wavefunctions. In numerous situations this solution is
good enough to obtain the cross sections results in comparison with
the experimental data.

The paper is organized as follows.  A Brief review of Multipole
matrix elements and reduced transition probabilities in sec.~{2}.
The relevant formalism and parameters, electric and magnetic
multipole matrix elements and the reduced transition probabilities
are defined in this section.  The findings of the model with
asymptotic wave functions are corroborated in more realistic
calculations using wave function generated from the Woods-Saxon
potentials and experimental data, in sec.~{3}. Summary and
conclusions follow in Section~{4}.

\section{Brief review of theoretical framework}

The computer code RADCAP calculates various quantities of interest
radiative capture reactions. The bound state wavefunctions of final
nuclei are given by $\Psi_{JM}\left(  \mathbf{r}\right)$ and the
ground-state wavefunction is normalized so that $ \int d^{3}r\
\left\vert \Psi_{JM}\left(\mathbf{r}\right)\right\vert ^{2}=1$.

The wavefunctions are calculated using the central($V_{0}(r)$),
spin-orbit($V_{S}(r)$) and the Coulomb potential($V_{C}(r)$)
potentials. The potentials $V_{0}(r)$ and $V_{S}(r)$ are given by
\begin{eqnarray}
&&V_{0}(r) =V_{0}\ f_{0}(r),\ V_{S}(r)=-\ V_{S0}\ \left(
{\frac{\hbar}{m_{\pi}c}}\right)  ^{2}\ {\frac{1}{r}}{\frac{d}{dr}}f_{S}(r)\nonumber\\
&&f_{i}(r)=\left[1+\exp\left(\frac{r-R_{i}}{a_{i}}\right)\right]^{-1}\,
\end{eqnarray}
where $V_{0}$, $V_{S0}$, $R_{0}$, $a_{0},$ $R_{S0}$, and $a_{S0}$
are adjusted so that the ground state energy $E_B$ or the energy of
an excited state, is reproduced.

The radial Schr\"{o}dinger equation for calculating of the
bound-state are given by solving
\end{multicols}
 \ruleup
\begin{equation}
-\frac{\hbar^{2}}{2m_{ab}}\left[\frac{d^{2}}{dr^{2}}-\frac{l\left(
l+1\right)  }{r^{2}}\right] u_{lj}^{J}\left(r\right)+\left[
V_{0}\left( r\right) +V_{C}\left(r\right)+\left\langle
\mathbf{s.l}\right\rangle \ V_{S0}\left(  r\right)  \right]
u_{lj}^{J}\left(  r\right)  =E_{i} u_{lj}^{J}\left(  r\right)
\label{bss}
\end{equation}
 \ruledown \vspace{0.5cm}
\begin{multicols}{2}

with $\left\langle \mathbf{s.l}\right\rangle =\left[
j(j+1)-l(l+1)-s(s+1)\right]  /2$.

The electric and magnetic dipole transitions are given by
introducing of the following operators~\cite{Bohr}
\begin{eqnarray}
\mathcal{O}_{E\lambda\mu}&=&e_{\lambda}\
r^{\lambda}Y_{\lambda\mu}\left( \widehat{\mathbf{r}}\right)  ,\\
\nonumber \mathcal{O}_{M1\mu}&=&\sqrt{\frac{3}{4\pi}}\mu_{N}\ \left[
e_{M}l_{\mu} +\sum_{i=a,b}g_{i}\left(  s_{i}\right)  _{\mu}\right]
\end{eqnarray}
where $e_{\lambda}=Z_{b}e\left(  -\frac{m_{a}}{m_{c}}\right)
^{\lambda}+Z_{a}e\left(  \frac{m_{b}}{m_{c}}\right)^{\lambda}$ and
$e_{M}=\left(
\frac{m_{a}^{2}Z_{a}}{m_{c}^{2}}+\frac{m_{b}^{2}Z_{b}}{m_{c}^{2}}\right)$
are the effective electric and magnetic charges, respectively.
$l_{\mu}$ and $s_{\mu}$ are the spherical components of order $\mu$
($\mu=-1,0,1$) of the orbital and spin angular momentum ($\mathbf{l=-}%
i\mathbf{r\times\nabla}$, and $\mathbf{s=\sigma}/2$) and $g_{i}$ are
the gyromagnetic factors of particles $a$ and $b$. The $\mu_{N}$ is
also nuclear magneton.

The matrix element for the transition $J_{0}M_{0}\longrightarrow JM$
is given by~\cite{Bohr,Edmonds}

\end{multicols}
 \ruleup
\begin{eqnarray}
&&\left\langle JM\left\vert \mathcal{O}_{E\lambda\mu}\right\vert
J_{0} M_{0}\right\rangle =\left\langle
J_{0}M_{0}\lambda\mu|JM\right\rangle \ \frac{\left\langle
J\left\Vert \mathcal{O}_{E\lambda}\right\Vert J_{0}\right\rangle
}{\sqrt{2J+1}},\\ \nonumber &&\left\langle J\left\Vert
\mathcal{O}_{E\lambda}\right\Vert J_{0}\right\rangle =\left(
-1\right)  ^{j+I_{a}+J_{0}+\lambda}\ \left[  \left(  2J+1\right)
\left(  2J_{0}+1\right)  \right] ^{1/2}\ \left\{
\begin{array}
[c]{ccc}%
j & J & I_{a}\\
J_{0} & j_{0} & \lambda
\end{array}
\right\}  \ \left\langle lj\left\Vert
\mathcal{O}_{E\lambda}\right\Vert l_{0}j_{0}\right\rangle _{J},
\label{joj0}
\end{eqnarray}
\ruledown \vspace{0.5cm}
\begin{multicols}{2}

where the subscript $J$ is a reminder that the matrix element is
spin dependent.  For $l_{0}+l+\lambda=\mathrm{odd}$, the reduced
matrix element is null and for $l_{0}+l+\lambda=\mathrm{even}$, is
given by
\end{multicols}
 \ruleup
\begin{equation}
\left\langle lj\left\Vert \mathcal{O}_{E\lambda}\right\Vert l_{0}
j_{0}\right\rangle _{J}=\frac{e_{\lambda}}{\sqrt{4\pi}}\ \left(
-1\right) ^{l_{0}+l+j_{0}-j}\
\frac{\hat{\lambda}\hat{j_{0}}}{\hat{\jmath}} \
<j_{0}\frac{1}{2}\lambda0|j\frac{1}{2}>\ \int_{0}^{\infty}dr\
r^{\lambda }\ u_{lj}^{J}\left(  r\right)  \
u_{l_{0}j_{0}}^{J_{0}}\left(  r\right)
.\label{lol0}%
\end{equation}
\ruledown \vspace{0.5cm}
\begin{multicols}{2}
At very low energies, the  transitions will be much smaller than the
electric transitions.  The M$_1$ contribution has to consider in the
cross sections for neutron photo-dissociation or radiative capture.
The M$_1$ transitions, in the case of sharp resonances, for the
$J=1^{+}$ state in $^{8}$B at $E_{R}=630$ keV above the proton
separation threshold play a role~\cite{Kim}.

The reduced matrix elements of M$_1$ transition, for $l\neq l_{0}$
the magnetic dipole matrix element is zero and for $l=l_{0}$, is
given by~\cite{Lawson}
\end{multicols}
 \ruleup
\begin{eqnarray}
&  \left\langle lj\left\Vert \mathcal{O}_{M1}\right\Vert l_{0}j_{0}
\right\rangle _{J}=\left(  -1\right) ^{j+I_{a}+J_{0}+1}\
\sqrt{\frac{3}{4\pi }}\ \widehat{J}\widehat{J}_{0}\left\{
\begin{array}
[c]{ccc}%
j & J & I_{a}\\
J_{0} & j_{0} & 1
\end{array}
\right\}  \mu_{N}\nonumber\\
&  \times\left\{  \frac{1}{\widehat{l}_{0}}e_{M}\left[
\frac{2\widetilde {j}_{0}}{\widehat{l}_{0}}\left(
l_{0}\delta_{j_{0},\ l_{0}+1/2}+\left( l_{0}+1\right)
\delta_{j_{0},\ l_{0}-1/2}\right)  +\left(  -1\right)
^{l_{0}+1/2-j}\frac{\widehat{j}_{0}}{\sqrt{2}}\delta_{j_{0},\
l_{0}\pm
1/2}\delta_{j,\ l_{0}\mp1/2}\right]  \right.  \nonumber\\
&  +g_{N}\frac{1}{\widehat{l}_{0}^{2}}\left[  \left(  -1\right)
^{l_{0}+1/2-j_{0}}\widetilde{j}_{0}\delta_{j,\ j_{0}}-\left(
-1\right)
^{l_{0}+1/2-j}\frac{\widehat{j}_{0}}{\sqrt{2}}\delta_{j_{0},\
l_{0}\pm
1/2}\delta_{j,\ l_{0}\mp1/2}\right]  \nonumber\\
&  \left.  +g_{a}\left(  -1\right) ^{I_{a}+j_{0}+J+1}\widehat{J}_{0}
\widehat{J}\widehat{I}_{a}\widetilde{I}_{a}\left\{
\begin{array}
[c]{ccc}%
I_{a} & J & j_{0}\\
J_{0} & I_{a} & 1
\end{array}
\right\}  \right\}  \int_{0}^{\infty}dr\ u_{lj}^{J}\left( r\right) \
u_{l_{0}j_{0}}^{J_{0}}\left(  r\right)  ,\label{ljolj}
\end{eqnarray}
\ruledown \vspace{0.5cm}
\begin{multicols}{2}
where $g_{N}$=5.586(-3.826) for the proton(neutron) and
$\mu_{a}=g_{a}\mu_{N}$ is the magnetic moment of the core nucleus.

The reduced transition probability $dB((E,B)\lambda) / dE$ of the
nucleus, $i$ into $j+k$, contains the information on the structure
in the initial ground state and the interaction in the final
continuum state. The reduced transition probability for a specific
electromagnetic transition $(E,B)\lambda$ to a final state with
momentum $\hbar k$ in the continuum is given by~\cite{Bertulani}
\end{multicols}
 \ruleup
\begin{eqnarray} \label{eq:dbde}
  \lefteqn{\frac{dB}{dE} ((E,B)\lambda, J_{i} s  \to k J_{f} s)
  =}\\ \nonumber &&\frac{2J_{f}+1}{2J_{i}+1}  \sum_{j_{f} l_{f}}
 \left| \sum_{j_{i} l_{i} j_{c}}
 \langle k J_{f}j_{f}l_{f}s j_{c} || {\mathcal M}((E,B)\lambda)
  || J_{i} j_{i} l_{i} s j_{c} \rangle \right|^{2}
 \frac{\mu k}{(2\pi)^{3}\hbar^{2}}
\end{eqnarray}
\ruledown \vspace{0.5cm}
\begin{multicols}{2}
The electric excitations ($E$) with multipole operator, is given by
\begin{equation}
 {\mathcal M}(E\lambda \mu) =
 Z_{\rm eff}^{(\lambda)}e  r^{\lambda} Y_{\lambda \mu}(\hat{r})
\end{equation}
where $Z_{\rm eff}^{(\lambda)} =
 Z_{b}\left(\frac{m_{c}}{m_{b}+m_{c}}\right)^{\lambda}
 +Z_{c}\left(-\frac{m_{b}}{m_{b}+m_{c}}\right)^{\lambda}$ is the
effective charge number.

For proton radiative capture the effective charge numbers for E$_1$
and E$_2$ have to consider both contributions in the cross sections
for Coulomb breakup, photo dissociation or radiative capture. In the
case of a neutron radiative capture, E$_1$ transition dominate the
low-lying electromagnetic strength and the E$_2$ contribution can be
neglected.

The initial and final state are given by the following wave
functions~\cite{Bertulani}

\end{multicols}
 \ruleup
\begin{eqnarray}
 \Phi_{i}(\vec{r})
 & = & \langle \vec{r} | J_{i} j_{i} l_{i} s j_{c} \rangle
 = \frac{1}{r}
 \sum_{m_{i}m_{c}} (j_{i} \: m_{i} \: j_{c} \: m_{c} | J_{i} \: M_{i} )
 f_{J_{i}j_{i}l_{i}}^{j_{c}}(r) {\mathcal Y}_{j_{i}m_{i}}^{l_{i}s}
(\hat{r}) \phi_{j_{c}m_{c}}\\ \nonumber
 \Phi_{f}(\vec{r})
 & = & \langle \vec{r} | \vec{k} J_{f} j_{f} l_{f} s j_{c} \rangle
 \\ \nonumber
 & = & \frac{4\pi}{kr}
 \sum_{m_{f}m_{c}} (j_{f} \: m_{f} \: j_{c} \: m_{c} | J_{f} \: M_{f} )
 g_{J_{f}j_{f}l_{f}}^{j_{c}}(r) i^{l_{f}}
 Y_{l_{f}m_{f}}^{\ast}(\hat{k})
 {\mathcal Y}_{j_{f}m_{f}}^{l_{f}s}
(\hat{r}) \phi_{j_{c}m_{c}} \: ,
\end{eqnarray}
\ruledown \vspace{0.5cm}
\begin{multicols}{2}
where $f^{j_{c}}_{J_{i}j_{i}l_{i}}(r)$ and
$g_{J_{f}j_{f}l_{f}}^{j_{c}}(r)$ are the radial wave functions and
$\phi_{j_{c}m_{c}}$ is also the wave function of the core. The
spinor spherical harmonics is denoted by ${\mathcal Y}_{jm}^{ls} =
\sum_{m_{l} m_{s}}
 ( l \: m_{l} \: s \: m_{s} | j \: m) Y_{lm}(\hat{r}) \chi_{s
 m_{s}}$.

 The reduced matrix element in (\ref{eq:dbde}) can be expressed
 as~\cite{Bertulani}
 \end{multicols}
 \ruleup
\begin{eqnarray} \label{eq:rme}
  \langle k J_{f} j_{f} l_{f} s j_{c} || {\mathcal M}(E\lambda)
 || J_{i} j_{i} l_{i} s j_{c}\rangle = \frac{4\pi Z_{\rm eff}^{(\lambda)}e}{k}
 D_{J_{i}j_{i}l_{i}}^{J_{f}j_{f}l_{f}}(\lambda s j_{c}) \:
 (-i)^{l_{f}} I_{J_{i}j_{i}l_{i}}^{J_{f}j_{f}l_{f}}(\lambda j_{c})
\end{eqnarray}
\ruledown \vspace{0.5cm}
\begin{multicols}{2}
where the angular momentum coupling coefficient
$D_{J_{i}j_{i}l_{i}}^{J_{f}j_{f}l_{f}}(\lambda s j_{c})$ and the
radial integral $I_{J_{i}j_{i}l_{i}}^{J_{f}j_{f}l_{f}}(\lambda
j_{c})$ are given by
\end{multicols}
 \ruleup
\begin{eqnarray}
D_{J_{i}j_{i}l_{i}}^{J_{f}j_{f}l_{f}}(\lambda s j_{c})  & = &
(-1)^{s+j_{i}+l_{f}+\lambda} (-1)^{j_{c}+J_{i}+j_{f}+\lambda}
 (l_{i} \: 0 \: \lambda \: 0 | l_{f} \: 0 )
 \sqrt{2j_{i}+1} \sqrt{2l_{i}+1}
 \sqrt{2J_{i}+1} \sqrt{2j_{f}+1}
\\ \nonumber & &
\times \sqrt{\frac{2\lambda+1}{4\pi}} \left\{ \begin{array}{ccc}
 l_{i} & s & j_{i} \\ j_{f} & \lambda & l_{f}
 \end{array} \right\}
 \left\{ \begin{array}{ccc}
 j_{i} & j_{c} & J_{i} \\ J_{f} & \lambda & j_{f}
 \end{array} \right\} \: ,\\ \nonumber
 I_{J_{i}j_{i}l_{i}}^{J_{f}j_{f}l_{f}}(\lambda j_{c})
 &=& \int_{0}^{\infty} dr \:
 g^{j_{c}\ast}_{J_{f}j_{f}l_{f}}(r)
 r^{\lambda} f^{j_{c}}_{J_{i}j_{i}l_{i}}(r)
\end{eqnarray}
\ruledown \vspace{0.5cm}
\begin{multicols}{2}
with the asymptotic radial wave functions for the bound state
\begin{equation} \label{eq:asymb}
 f^{j_{c}}_{J_{i}j_{i}l_{i}}(r)
 \to C^{j_{c}}_{J_{i}j_{i}l_{i}} W_{-\eta_{i}, l_{i}+1/2}
 (2qr) ,
\end{equation}
and the asymptotic form of the continuum state for the scattering
state
\end{multicols}
 \ruleup
\begin{eqnarray} \label{eq:asyms}
 g^{j_{c}}_{J_{f}j_{f}l_{f}}(r) & \to &
  \exp\left[i(\sigma_{l_{f}}+\delta^{j_{c}}_{J_{f}j_{f}l_{f}})\right]
 \times
 \left[ \cos (\delta^{j_{c}}_{J_{f}j_{f}l_{f}}) \: F_{l_{f}}(\eta_{f};kr)
 + \sin ( \delta^{j_{c}}_{J_{f}j_{f}l_{f}}) \:  G_{l_{f}}(\eta_{f};kr) \right]
\end{eqnarray}
\ruledown \vspace{0.5cm}
\begin{multicols}{2}
where $C^{j_{c}}_{J_{i}j_{i}l_{i}}$, $W_{-\eta_{i}, l_{i}+1/2}$,
$F_{l_{f}}$, $G_{l_{f}}$ and $\eta_{f}=\eta_{i}/x$ are the
asymptotic normalization coefficient, Whittaker function, regular
Coulomb wave functions, irregular Coulomb wave functions and the
Sommerfeld parameter, respectively~\cite{Abramowitz}.

Cross-section for non-same particles without spin, is defined as
follows:
\begin{eqnarray}
\sigma_{(E,B)l}^{cap}(\varepsilon)=\frac{\pi
\hbar^2}{2\mu\varepsilon}(2l+1)T_{(E,B)l}\,,
\end{eqnarray}
where $T_{(E,B)l}$ is the transition probability. Finally, the total
cross section for a transition is arbitrary:
\begin{eqnarray}
 \sigma^{cap}(\varepsilon)=\sum_{l}(\sigma_{El}^{cap}(\varepsilon)+\sigma_{Ml}^{cap}(\varepsilon))
\end{eqnarray}
The total cross section for an arbitrary transition is defined as:
\begin{eqnarray}
 \sigma^{cap}(\varepsilon)=S(\varepsilon)\frac{1}{\varepsilon}e^{-2\pi\eta}
\end{eqnarray}
In this equation$ S(\varepsilon)$ astrophysical factor and $
\eta=\frac{Z_c Z_\alpha e^2}{\hbar}\sqrt{\frac{\mu}{2\varepsilon}}$
are Parameters Samrfyld. We use the astrophysical S-factor because
it is a well-define function with little changes and it is easier to
analyze.

\section{Result and conclusions}

The potential model and the RADCAP computer code are proper
theoretical frameworks to describe the ground state properties of
$^{16}$O for the reaction $^{12}$C($\alpha$,$\gamma$)$^{16}$O. To
evaluate the radiation capture reaction
$^{12}$C($\alpha$,$\gamma$)$^{16}$O, Schrodinger equation using
Wood-Saxon potential and with solved specific parameters and bound
continuum states of the reaction is obtained with very good
accuracy, and finally using the formulation of the second part of
the paper, the astrophysical S-factor is calculated for transition
E$_2$.
\end{multicols}
 \ruleup
\begin{center}
\tabcaption{ \label{tab1} The set of Woods-Saxon potential
parameters, applied for calculation. } \footnotesize
\begin{tabular}{ccccccc}
\hline\hline
 $V_0$(MeV) & $R_0$ (fm) & $a_0$ (fm) & $V_{S0}(MeV)$ & $R_{S0}$ (fm)& $a_{S0}$ (fm) & $R_C$ (fm)\\
 \hline
   -51.8 & 2.41 & 0.644 & 39.54 & 2.291 & 0.644 & 2.41 \\
\hline
\end{tabular}
\end{center}
\vspace{0.5cm}
\begin{multicols}{2}

\end{multicols}
 \ruleup
 \begin{center}
\includegraphics*[width=16cm]{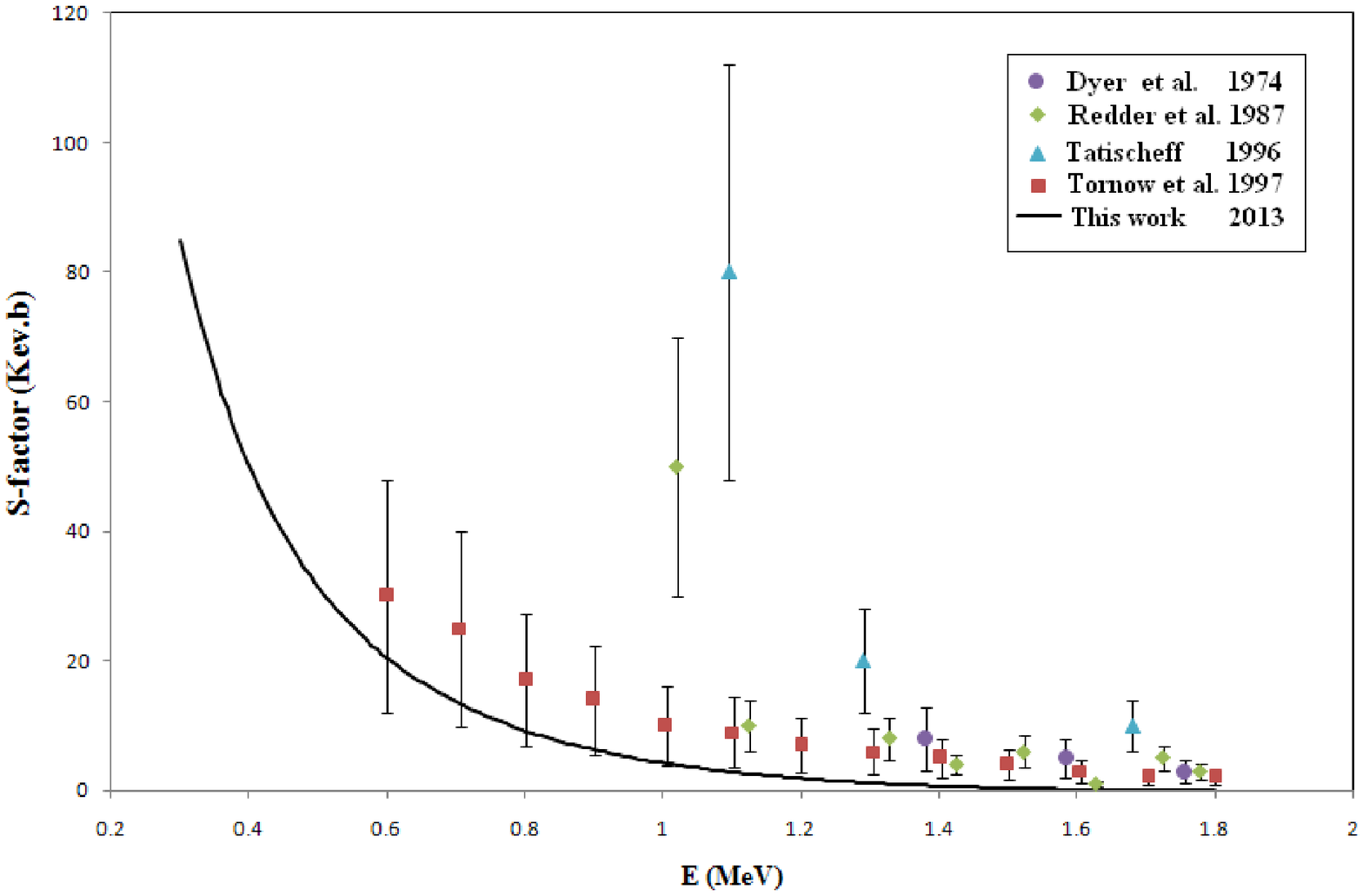}
\figcaption{\label{fig2}The astrophysical $S$-Factor for the
$^{12}$C($\alpha$,$\gamma$)$^{16}$O reaction. The calculated results
are given by the solid line and the available experimental
data~\cite{Tatischeff,Redder,Dyer,Tornow} are shown as different
symbols.}
\end{center}
\ruledown \vspace{0.5cm}
\begin{multicols}{2}
The set of Woods-Saxon potential parameters, applied for calculation
are given in table~{1}. The results for the astrophysical S-Factor
of $^4$He--$^{12}$C radiative capture process is presented in
Fig.~1, along with the experimental
data~\cite{Tatischeff,Redder,Dyer}, at solar energies 0.1--2 MeV.
\begin{center}
\tabcaption{ \label{tab1} Some evaluated astrophysical S(E$_cm$=300
keV)--factor value for experimental data for the
$^{12}$C($\alpha$,$\gamma$)$^{16}$O reaction in keV.b }
\footnotesize
\begin{tabular}{cccc}
\hline\hline
 Reference & year  & S(E$_2$)  \\
 \hline
Sch\"{u}rmann et al.~\cite{Schurmann} & 2012 & 73.4  \\
Oulebsir et al.~\cite{Oulebsir} & 2012 & 50 $\pm$ 19  \\
Hammer et al.~\cite{Hammer}& 2005 & 81 $\pm$ 22 \\
Kunz et al.~\cite{Kunz} & 2001 & 85 $\pm$ 30  \\
Redder et al.~\cite{Redder} & 1987 & 80 $\pm$ 25  \\
This work & 2013 & 84.97  \\
 \hline
\end{tabular}
\end{center}

The value of this quantity is obtained at 300 keV transition energy
E$_2$ is found to be 84.97 keV.b which is reasonably agreement with
some evaluated value for experimental data that shown in table~{2}.
Here, no significant difference has been seen between the results
obtained with the present model based on the potential model, and
some evaluated value for experimental data in papers with potential
model.  In the other theoretical approach by using the cluster model
the astrophysical S-Factor of $^4$He--$^{12}$C have been calculated
to be S(300)=16.0 keV.b~\cite{Dubovichenko2}.

\section{Summary and conclusions}

Radioactive capture $^{12}$C($\alpha$,$\gamma$)$^{16}$O  is one of
the most important reactions in nuclear astrophysics. The reaction
amount determines the relative abundance of most elements in red
giant stars, neutron stars and black holes. In general, the electric
dipole radiation E$_1$ are much stronger than quadrupole radiation
from electric E$_2$. And electric dipole transitions between states
with the same isospin forbidden in the first order. Because state
1$^+$ and 0$^+$ ground state Nuclei $^{16}$O with are isospin T = 0,
Thus the electric dipole radiations are not at the first order
between two levels and electric dipole radiation will be the second
order and Electric dipole radiation is same order with the electric
quadrupole radiation. Therefore, we must consider the effects of
both radiations. In comparison with other theoretical methods and
available experimental data, good agreement is achieved for the
astrophysical S-factor of this process.

\section{Acknowledgements} The authors would like to acknowledge C.
A. Bertulani, for online RADCAP computer code.

\end{multicols}

\clearpage

\end{document}